\documentclass[pdflatex,sn-mathphys]{sn-jnl}
\usepackage{fix-cm}

\usepackage{fix-cm}
\usepackage{float}
\usepackage{endnotes}
\let\footnote=\endnote
\usepackage{colortbl}
\newcolumntype{L}[1]{@{}>{\raggedright\arraybackslash}p{#1}} 
\usepackage{xcolor}
\definecolor{darkblue}{rgb}{0.0, 0.0, 0.55} 

\jyear{2021}%

\theoremstyle{thmstyleone}%
\usepackage{longtable}
\usepackage{array}
\theoremstyle{thmstyletwo}%

\theoremstyle{thmstylethree}%

\begin{document}

\title[Indicators for monitoring a National Artificial Intelligence Strategy]{Identifying relevant indicators for monitoring a National Artificial Intelligence Strategy.}


\author*[1]{\fnm{Renata} \sur{Pelissari}}\email{renata.infante@mackenzie.com}

\author[2]{\fnm{Ricardo} \sur{Suyama}}\email{ricardo.suyama@ufabc.edu.br}
\equalcont{These authors contributed equally to this work.}

\author[3]{\fnm{Leonardo} \sur{Tomazeli Duarte}}\email{leonardo.duarte@fca.unicamp.br}
\equalcont{These authors contributed equally to this work.}

\author[4]{\fnm{Henrique} \sur{Sá Earp}}\email{henrique.saearp@ime.unicamp.br}
\equalcont{These authors contributed equally to this work.}

\affil[1]{\orgdiv{School of Engineering}, \orgname{Mackenzie Presbyterian University}, \orgaddress{\street{Rua da Consolação, 930}, \city{São Paulo}, \postcode{01302-907}, \state{SP}, \country{Brazil}}}

\affil[2]{\orgdiv{Center for Engineering, Modeling and Applied Social Sciences (CECS)}, \orgname{Federal University of ABC (UFABC)}, \orgaddress{\street{Avenida dos Estados, s/n}, \city{Santo André}, \postcode{09210-580}, \state{SP}, \country{Brazil}}}

\affil[3]{\orgdiv{School of Applied Sciences}, \orgname{UNICAMP}, \orgaddress{\street{Rua Pedro Zaccaria, 1300}, \city{Limeira}, \postcode{13484-350}, \state{SP}, \country{Brazil}}}

\affil[4]{\orgdiv{Institute of Mathematics, Statistics, and Scientific Computation (IMECC)}, \orgname{UNICAMP}, \orgaddress{\street{Rua Sérgio Buarque de Holanda, 651}, \city{Campinas}, \postcode{13083-859}, \state{SP}, \country{Brazil}}}

\abstract{How can a National Artificial Intelligence Strategy be effectively monitored? To address this question, we propose a methodology consisting of two key components. First, it involves identifying relevant indicators within national AI strategies. Second, it assesses the alignment between these indicators and the strategic actions of a specific government’s AI strategy, allowing for a critical evaluation of its monitoring measures. In addition, identifying these indicators helps to assess the overall quality of the strategy structure. A lack of alignment between strategic actions and identified indicators can reveal gaps or blind spots in the strategy. This methodology is demonstrated using the Brazilian AI strategy as a case study.
 }




\maketitle


\newpage
\section{Introduction}

Artificial intelligence (AI) has been one of the main drivers for the development of cutting-edge technologies that are impacting society at different levels \cite{MAKRIDAKIS201746, schwab2017fourth, bessen2018ai}. To harness the benefits of AI, while mitigating the risks, governments are developing National Strategies, seeking geopolitical protagonism and leveraging economic, social and cultural progress \cite{European_Comission_Strategies2021}. Launched in 2017, the Pan-Canadian Artificial Intelligence Strategy \cite{CAN_AI_Strategy} was the first national strategy with the goal of guiding the priorities of AI policy at the country level \cite{Radu2021}. Finland also developed its national AI strategy in 2017, closely followed by Japan, France, Germany, and the United Kingdom in 2018. More than 30 other countries and regions have launched their AI strategies as of 2021\footnote{\url{https://hai.stanford.edu/news/state-ai-10-charts.} Accessed date: 2025-01-23}, including Brazil.

The formulation of long-term plans such as National Artificial Intelligence Strategies provides a comprehensive perspective on government initiatives aimed at social and economic development over periods of five years or more. A strategic plan should ideally articulate how a nation envisions the opportunities presented by AI, considering the strengths and weaknesses of the country  \cite{berryhill2019hello}. This vision should guide capability-building efforts, including targeted investments across various sectors and industries, as well as the development of regulatory frameworks and governance protocols to mitigate AI-related risks \cite{berryhill2019hello, Wirtz2019}. Thus, strategic plans in the public sector serve as an essential roadmap, outlining national priorities and the pathways to achieve them. Although not flawless, these plans can mirror current perspectives of the countries on how to engage with the advances regarding AI \cite{Wirtz2021, halterlein2024ai}. Recently, artificial intelligence strategies at various levels of government have been under development and are expected to receive increased attention in the coming years, given that AI is intricately linked to numerous policy areas where responsibilities are already distributed between national and subnational levels \cite{liebig2022subnational}.

Usually, a National Artificial Intelligence Strategy begins by providing a diagnosis of the country's situation regarding AI, detailing current capabilities, infrastructural strengths, challenges, and potential avenues for growth \cite{berryhill2019hello}. This analysis serves as the foundation for defining strategies and action plans across various dimensions, such as Research and Innovation, Education and Workforce Development, Digital Infrastructure, Ethics and Regulation, Sectoral Applications, Social Inclusion and Diversity, Security and Defense, and International Collaboration \cite{FATIMA2020178, berryhill2019hello, Wirtz2019}. Despite the similarity in the dimensions considered in the strategies, nations have pursued distinct priorities, such as academic excellence in Canada, skills development in South Korea, technological sovereignty in Germany, and education and AI services for public administration in Finland \cite{Radu2021}.

Most strategies contain aspirational plans and do not elaborate on the actual implementation roadmaps or tracking mechanisms \cite{FATIMA2020178, berryhill2019hello}. These mechanisms, once defined and disclosed, are typically made available in documents updating the original plan (as presented by Germany \cite{DEU_Strategy_Update_2020} and France \cite{FRA_AI_Strategy_Phase2}), in monitoring reports (as published by Canada \cite{CAN_AI_Report}), or on AI monitoring platforms (such as the Brazilian AI Observatory\footnote{\url{https://www.gov.br/mcti/pt-br/acompanhe-o-mcti/transformacaodigital/arquivosinteligenciaartificial/1_ebia-reuniao-ro_7_24_05_2023_anexo_2_eixo2-pdf.pdf}. Accessed: 2025-01-23}), as discussed in more depth in Section \ref{sec:monitoring}.
Whatever monitoring mechanism is adopted, the definition of viable and relevant performance indicators for a national strategy, capable of reliably representing the different impacts of AI, has proven to be one of the main actions among leading countries in this field \cite{European_Comission_Strategies2021}.
The task of defining indicators requires an understanding of aspects ranging, e.g., from microeconomic impacts to the real impact on the population's quality of life \cite{Julnes2014, Neely1995, Neely2007}. Furthermore, the indicators defined must be easy to acquire and to update periodically, through feasible and previously established processes \cite{OECD2020}. 

Given that context, this article presents a methodology to identify relevant indicators for monitoring a National Artificial Intelligence Strategy (NAIS), with the goal of establishing best practices for performance assessment, based on the authors' recent experience in providing subsidies for the revision\footnote{\url{https://www.gov.br/mcti/pt-br/acompanhe-o-mcti/noticias/2024/07/plano-brasileiro-de-ia-tera-supercomputador-e-investimento-de-r-23-bilhoes-em-quatro-anos/ia_para_o_bem_de_todos.pdf/view}. Accessed: 2024-08-21.} of the Brazilian Artificial Intelligence Strategy (in Portuguese, \textit{Estratégia Brasileira de Inteligência Artificial} - EBIA) \cite{EBIA}. This methodology addresses the following research questions:
 \begin{itemize}
    \item RQ1: What are the most relevant strategic indicators for monitoring a National Artificial Intelligence Strategy?
    \item RQ2: What are the correspondences between these indicators and the structure of the analyzed National Artificial Intelligence Strategy?
    \item RQ3: Which feedback can be obtained, either about the monitoring process or the strategy itself, based on the results of these correspondences?
 \end{itemize}
 
While the outcomes we obtain are thematically centered on the Brazilian perspective, we seek to present the proposed methodology in a systematic way, amenable to other national experiences, while illustrating its application within the context of EBIA. 

The paper is structured as follows. Section \ref{sec:monitoring} provides a brief background on the mechanisms used for monitoring a NAIS. Our proposed methodology to identify relevant performance indicators  is presented in Section \ref{sec:ProposedMethodology}, and it is applied in the context of EBIA in Section \ref{sec:EBIA}. We conclude briefly in Section \ref{sec:Conclusion}.

\section{Monitoring National AI Strategies}
\label{sec:monitoring}

Once an AI strategy is defined, it is essential to monitor its progress and measure the impact of its programs and initiatives to enable continuous improvement. Regardless of the specific monitoring and control mechanisms employed by various countries and commissions worldwide, these mechanisms must be comprehensive enough to track and assess all aspects of the AI plan in question. As highlighted by the European Commission in its 2021 review of the Coordinated Plan on AI \cite{eu2021coordinated}, Member States are encouraged to develop and promote instruments that facilitate regular monitoring, coordination, and evaluation of AI policy actions. Likewise, as noted in \cite{European_Comission_Strategies2021}, the Member States of the European Commission unanimously agree on the importance of evaluating both the progress of strategy implementation and the outcomes of their policy actions.

To this end, various mechanisms exist. As a reference, we can mention reports provided by some countries that disclose indicators identifying the level of development of the previously proposed AI strategy. Canada, after proposing its AI strategy in 2017, released its first progress report in 2023, titled ``AICan: The Impact of the Pan-Canadian AI Strategy'' \cite{CAN_AI_Report}. Similarly, in November 2018, the German Federal Government launched its National AI strategy and, in 2020, released an update that also contains an analysis of the initial strategy \cite{DEU_Strategy_Update_2020}. Like Germany and Canada, other countries that have released progress reports on their national strategies include Australia \cite{AUS_AI_Strategy} and France \cite{FRA_AI_Strategy_Phase2} and the United Kingdom \cite{GBR_AI_Strategy_Report}.

Some countries, including Brazil, have proposed creating AI observatories to monitor the implementation of their AI strategies, among other tasks. The Brazilian Observatory of Artificial Intelligence (OBIA), launched in 2024, is an initiative aimed at monitoring, analyzing, and promoting the development and application of AI in Brazil. Its goal is to provide information, data, and analyses about the state of AI in the country while fostering collaboration among key stakeholders, including government, academia, the private sector, and civil society. Spain also hosts AI observatories, such as the \emph{Observatorio de Inteligencia Artificial para la Administración Pública} in the Valencian Community\footnote{\url{https://observatorioia.gva.es/es/}} and the Observatorio de IA en Salud in Catalonia\footnote{\url{https://iasalut.cat/es/observatori-ia-en-salut/observatori-dintelligencia-artificial-en-salut/}. Last accessed on August 20, 2024}.

The European Commission has launched the AI Watch initiative, which monitors AI development, adoption, and impact across the continent. AI Watch provides stakeholders with consistent, accessible, and up-to-date information on various AI-related trends, including research and development, jobs, skills, and the latest AI news. It also features a section called ``Live Data,'' which uses real-time data to display trends in AI development, application, and sectoral use \cite{eu2021coordinated}. The OECD has similarly created the AI Policy Observatory, which aims to monitor global AI policies \cite{OECD2020}.
Despite these efforts, few countries have established key performance indicators (KPIs) or milestones in their AI strategies for the coming years. Within the European Commission, only Finland, France, and Hungary have outlined KPIs to evaluate policy outcomes \cite{European_Comission_Strategies2021}. Outside Europe, the situation is similar, with only a few countries, such as Canada and South Korea, including this information in their strategies.

\section{Methodology}
\label{sec:ProposedMethodology}

We present the methodology for defining indicators for monitoring National Artificial Intelligence Strategies. It is structured into two main stages, which in turn include various activities, as discussed below.

\subsection{Stage 1: Identification of strategic and relevant AI indicators}

In this stage, the objective is to identify the most relevant indicators used to monitor AI strategies.
The starting point is a predefined list of indicators, gathered from readily available databases such as socioeconomic census data and other government agency reports related to science, technology, and innovation. The indicators from this preliminary list are then compared with those found in pre-selected national AI strategies, chosen according to criteria of geopolitical relevance to the country in question. For instance, Brazil might wish to compare its indicators with other Latin American economies, such as Argentina or Mexico, to other advanced developing countries, or indeed BRICS nations, as well as international benchmark countries like some or all of the G7 nations. Another simple but important consideration is the emphasis and robustness of use of indicators by the mentioned strategies, since some may be more keen than others to draw and justify their conclusions from quantitative data. 

Considering the indicators found in the national strategies that stand out, we identify relevant ones, which were not present in the preliminary list. The selection of new indicators is guided by their frequency in the surveyed documents and their effective role in shaping situational analyses and incentive policies. 

The stage 1 process involves the following activities:

\begin{itemize}
    \item \textbf{Activity 1.1:} Definition of a preliminary list of indicators. This list may include those already highlighted in the strategy of the analysed country. A taxonomy shall be defined to categorise these indicators.
    \item \textbf{Activity 1.2:} Selection of a set of countries for the study, based on the following criteria: (i) global leadership in AI; (ii) continental and/or regional representativeness; (iii) development stage compatible with the scenario of the analysed country. Other criteria can be defined and used without compromising the proposed methodology.
    \item \textbf{Activity 1.3:} Retrieval of National AI Strategies, or equivalent documents such as plans and policies, of all sample countries. We suggest starting with the OECD thematic portal \cite{oecd_portal_ai_2021}.
    \item \textbf{Activity 1.4:} Identification of standout National AI Strategies in relation to the adoption of indicators. We propose to consider as standout National AI Strategies those with a number of indicators greater than a predefined threshold.
    \item \textbf{Activity 1.5:} Identification of the presence and analysis of the prevalence of the initial list of indicators in the National AI Strategies. From the documents gathered in Activity 1.3, it is possible to classify these indicators according to their prevalence among the selected countries as a preliminary metric of their international strategic relevance. For each indicator $i$, we compute its frequency $f_i$ among the sample countries and determine the average frequency $\bar{f} $ and the standard deviation $\sigma$ of the series. Indicators are then categorized as follows:
    \begin{center}
\begin{tabular}{lc}
    \textbf{Frequency}        & \textbf{Condition} \\
    Highly Prevalent         & $f_i \geq \bar{f}  + \sigma$ \\
    Prevalent                & $\max\{\bar{f}  - \sigma,0\} < f_i < \bar{f}  + \sigma$ \\
    Irrelevant               & $f_i \leq \max\{\bar{f}  - \sigma,0\}$ \\
\end{tabular}
\end{center}
    Note that in highly dispersed series, the difference $\bar{f}  - \sigma$ may be negative, in which case we truncate the frequency difference at zero.
    
    \item \textbf{Activity 1.6:} Based on the in-depth analysis of standout National AI Strategies, we shall identify relevant indicators that were not considered in the preliminary list. Then, the identified indicators are categorized according to the taxonomy of indicators adopted in the preliminary list, grouping similar indicators found in more than one strategy and proposing new dimensions, areas, and units of analysis when necessary.
    \item \textbf{Activity 1.7:} We formulate elements for the feasibility study of collecting the proposed indicators. This information aims to support the final decision on which indicators to effectively monitor from the set of proposed indicators.
\end{itemize}
    
The expected outcome from Stage 1 is a consolidated set of feasible indicators. This set encompasses the indicators initially proposed by the preliminary list and highly prevalent in a broad scope of the analyzed strategies, as well as those new indicators identified in the standout strategies. In Section \ref{subsec:Stage1br}, we describe the application of this methodology to define feasible indicators for the Brazilian Artificial Intelligence Strategy. In this case, starting from an initial list of 56 indicators and analyzing strategies from 13 countries, we identified 30 feasible indicators spanning five dimensions of analysis.

\subsection{Stage 2: Correspondence between relevant indicators and the structure of the analyzed strategy}

In this stage, the objective is to relate the set of indicators identified in Stage 1 to the thematic structure of the national AI strategy being analyzed. The activities in Stage 2 are organized as follows:

\begin{itemize}
    \item \textbf{Activity 2.1:} Identification of the thematic structure of the strategy. This involves determining the core themes, axes, or dimensions that define the strategy’s key objectives. These elements act as the guiding framework, shaping the overall direction and focus of the strategy.
    
    \item \textbf{Activity 2.2:} Development of a quantitative tool to assess the degree of correspondence between the consolidated set of indicators and the thematic structure of the analyzed strategy. This tool will quantify how well the indicators reflect the key components, axes, and dimensions outlined in the strategy, enabling accurate monitoring and evaluation of progress.
    
    \item \textbf{Activity 2.3:} Definition of visualization tools that effectively display the correspondences identified in Activity 2.2. These tools will be designed to present data in a clear and intuitive manner, facilitating interpretation and analysis of the results.
    
    \item \textbf{Activity 2.4:} Application of the quantitative and visualization tools to produce materials that illustrate the levels of correspondence between the indicators and the strategy's thematic structure.
\end{itemize}

The outputs of Stage 2 are a correspondence map outlining the alignment between the set of indicators and the strategy's thematic structure, a frequency analysis of how often each indicator is present within the various dimensions of the strategy, and a list of elements to consider for the feasibility study regarding the collection of the proposed indicators.

In the Brazilian case, discussed in Section \ref{subsec:Stage2br}, we adopt a quantitative tool based on a binary correspondence between the indicator and the thematic structure of the strategy. Despite its simplicity, this approach not only facilitates a straightforward mapping of indicators to strategy themes but also aids in identifying gaps where certain themes may lack adequate measurement.

\subsection{Stage 3: Identifying Patterns}

In this stage, the goal is to highlight patterns identified in the correspondence analysis from the previous stage. The results obtained here allow for the selection of indicators to be incorporated into the strategy. Additionally, potential improvements to the structure of the analyzed strategy can be identified. Although structural differences between national strategies are influenced by varying scientific, technological, and economic priorities, a lack of correspondence between highly relevant indicators and the strategy’s framework may indicate a blind spot in the analyzed strategy.

For instance, in the Brazilian case, as discussed in Section \ref{subsec:Stage3br}, this method exposed specific gaps where key indicators prevalent in national strategies were not well-aligned with the existing thematic structure. As a result of this analysis, several potential improvements to the strategy were proposed and are detailed in Section \ref{subsec:ImproveEBIA}.

\section{Application: The Brazilian Artificial Intelligence Strategy} \label{sec:EBIA}

The Brazilian Artificial Intelligence Strategy (\textit{Estratégia Brasileira de Inteligência Artificial} - EBIA) \cite{EBIA} is a comprehensive document developed by the Ministry of Science, Technology, and Innovations of Brazil, aimed at fostering and regulating Artificial Intelligence (AI) within the country. Its goal is to promote the ethical and responsible development of AI, maximizing its social and economic benefits while addressing potential risks and challenges.

\begin{figure}[!ht]
\centering
\includegraphics[width=0.8\textwidth]{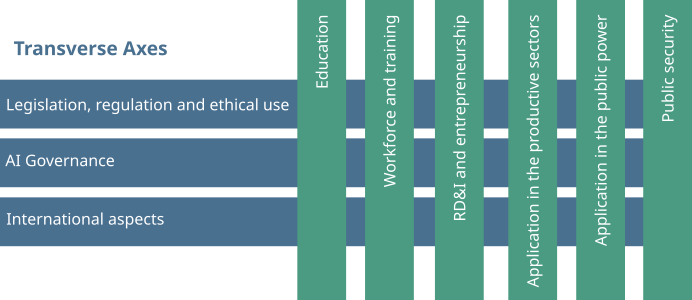}
\caption{Structure of EBIA and its nine thematic axes.}\label{Fig:Eixos_EBIA}
\end{figure}

EBIA adopts a holistic approach, structured around three cross-cutting axes and six vertical axes, forming an ideal $3 \times 6$ matrix, as shown in Figure \ref{Fig:Eixos_EBIA}. This structure defines a total of nine thematic axes, each tackling critical aspects for AI development in Brazil, as summarized in Table \ref{Tab:Eixos_EBIA}. The specific strategic actions for each axis are detailed in Table \ref{tab:actionsEBIA} in Appendix A.

EBIA is described as a ``living’’ document, designed to undergo regular updates and adjustments in response to the rapid evolution of AI technologies. Its objective is to establish Brazil as an important player on the global technology stage, enhancing quality of life for its citizens and promoting sustainable development.

\begin{table}[!ht]
    \centering
    \small
    \begin{tabular}{|p{0.4\textwidth}|p{0.5\textwidth}|}
    \hline
    \textbf{Thematic axis} & \textbf{Description} \\
    \hline
    Legislation, Regulation and Ethical Use (LRE) & Focuses on creating a legal and regulatory framework to ensure ethical development of AI. \\\hline
    Governance of AI (GIA) & Proposes governance structures to ensure the application of ethical principles in AI. \\\hline
    International Aspects (AI) & Highlights the importance of international collaboration in AI development. \\\hline
    Education (ED) & Addresses the need to qualify the population for future demands of the labor market impacted by AI. \\\hline
    Workforce and Training (FTC) & Emphasizes the development of skills necessary to adapt to digital transformation. \\\hline
    Research, Development, Innovation and Entrepreneurship (R\&D) & Encourages research and development in AI to promote innovation and entrepreneurship. \\\hline
    Application in Productive Sectors (ASP) & Examines how AI can be applied to increase efficiency and innovation in productive sectors. \\\hline
    Application in Public Power (APP) & Discusses the use of AI to improve the delivery of public services. \\\hline
    Public Security (SP) & Explores the use of AI to enhance public security through data analysis and automation. \\
    \hline
    \end{tabular}
    \caption{Description of the nine thematic axes of EBIA.}\label{Tab:Eixos_EBIA}
\end{table}

Our aim in this section is to apply the proposed framework in order to identify a list of indicators that could be used for monitoring EBIA.

\subsection{Stage 1}\label{subsec:Stage1br}

In the present case study, Activity 1.1 relies on a preliminary list of indicators provided by the Center for Management and Strategic Studies (CGEE), a Brazilian social organization affiliated with the Ministry of Science, Technology, and Innovation. Established in 2001, CGEE’s mission is to conduct forward-looking studies and research in science and technology, as well as to evaluate the economic and social impacts of scientific and technological policies, programs, and projects. The preliminary list from CGEE consists of 56 indicators grouped into seven dimensions, which we categorize alphabetically as follows:
\begin{itemize}
\item[A.] Adoption and use of AI-based applications
\item[B.] Knowledge production
\item[C.] Training
\item[D.] Skills and employment
\item[E.] Monitoring of international agendas
\item[F.] Monitoring of political debates and regulatory aspects
\item[G.] Monitoring of trends and innovation
\end{itemize}

Each indicator is assigned a unique alphanumeric code, ranging from A1 to G1, to simplify internal referencing. The complete list is presented in Table \ref{tab_lista_preliminar} of Appendix A.

Next, we select a set of countries for the study. We chose 13 countries based on the criteria outlined in Activity 1.2, aiming for a representative sample: South Africa, Germany, Argentina, Australia, Canada, Chile, China, South Korea, USA, France, India, Mexico, and the United Kingdom. The selection includes countries from all continents, three Latin American nations (including one Mercosur member), two European Union members (plus the UK), five G7 countries, and three BRICS members.

After selecting the countries, Activity 1.3 involves searching for the National Artificial Intelligence Strategies or equivalent documents of the selected countries, using the OECD thematic portal \cite{oecd_portal_ai_2021} as a starting point. Only South Africa lacks a formal document for its National AI Strategy, and thus it is excluded from further analysis. Table \ref{Tab:analise_documentos} provides a brief qualitative analysis of these documents, contextualizing each nation’s approach to systematically collecting AI indicators.

\begin{longtable}{|p{0.15\textwidth}|p{0.75\textwidth}|}
    \caption{Comments on National Strategies for Artificial Intelligence.}\label{Tab:analise_documentos} \\
    \hline 
    \small
    \textbf{Country} & \textbf{Observations} \\
    \hline
    Argentina & Argentina's strategy (\textit{Plan Nacional de Inteligencia Artificial}) \cite{ARG_AI_Strategy}, published in 2019, presents diagnoses in different axes, proposing 43 specific objectives and 79 actions. More than 70 indicators are also defined, distributed in the axes defined in the document. However, there is no indication of the source of the data for the indicators. \\
    \hline
    Australia & The plan released by the Australian government in 2021 (Artificial Intelligence Action Plan) \cite{AUS_AI_Strategy} brings a series of actions in four focal areas related to the use of AI in business, development and attraction of AI talents, use of AI to solve national challenges, and responsible and inclusive use of AI. It is noteworthy that the strategy presents short-term measures, with budgets and schedules for their implementation, but does not clearly mention indicators to be monitored. \\
    \hline
    Canada & Canada has an Artificial Intelligence strategy called (Pan-Canadian Artificial Intelligence Strategy) \cite{CAN_AI_Strategy}, launched in 2017, which seeks to position the country as a world leader in AI research and innovation. The strategy is based on three pillars: research, talents, and adoption of AI in key sectors of the economy. The Canadian government has invested in centers of excellence in research at various universities in the country, promoting collaboration between academia, industry, and government. Additionally, there is an emphasis on innovation in AI startups and attracting talent to boost AI development. \\
    \hline
    Chile & Chile has shown interest in promoting AI development, with initiatives such as the creation of the National Center for Artificial Intelligence (CENIA), which aims to boost AI research, development, and application in various sectors. Additionally, the country has implemented the National Digital Transformation Plan (Política Nacional de Inteligencia Artificial) \cite{CHL_AI_Strategy}, which includes AI as one of the strategic areas to drive digital transformation. It is important to highlight that Chile's strategy is marked by ethical pillars, principles, and social values, but it has few specific indicators, most of which are of international origin and none specific to Chile. \\
    \hline
    China & China launched the ``Next Generation Artificial Intelligence Development Plan'' \cite{CHN_AI_Strategy} in 2017, with the goal of becoming a world leader in AI by 2030. The Chinese strategy covers various areas such as research, innovation, talent development, regulation, ethics, and the application of AI in sectors such as health, transportation, industry, and agriculture. The Chinese government has been heavily investing in AI research and development, as well as in startups and technology companies. The Chinese document does not mention any quantitative metrics or indicators. \\
    \hline
    France & Although it does not present indicators, the French strategy (AI for Humanity)\cite{FRA_AI_Strategy}, based on a report published in 2018 \cite{FRA_AI_Diagnostic}, presents a series of initiatives for which performance indicators can be envisioned. Thus, the correspondences made in our analyses considered the mentioned initiatives as a basis for comparison with the indicators initially listed by CGEE. \\
    \hline
    Germany & Published the Artificial Intelligence strategy in 2018 (\textit{Strategie Künstliche Intelligenz der Bundesregierung}) \cite{DEU_AI_Strategy}, and later published an update of the document in 2020 (Strategie Künstliche Intelligenz der Bundesregierung: Fortschreibung 2020) \cite{DEU_Strategy_Update_2020}. It has at least two official platforms with information related to the Artificial Intelligence strategy: the first, supported by the Ministry of Education and Research, called Lernende Systeme; and the second, the national observatory for Artificial Intelligence in Work and Society, under the responsibility of the Ministry of Labor and Social Affairs. 
    \\
    \hline
    India & The Indian strategy (National Strategy for Artificial Intelligence) \cite{IND_AI_Strategy}, published in 2018, presents challenges and lines of action on different fronts but does not include performance indicators. The correspondence was made considering actions for which indicators can be easily envisioned. \\
    \hline
    Mexico & The Mexican strategy (\textit{Agenda Nacional Mexicana de Inteligencia Artificial}) \cite{MEX_AI_Strategy}, published in 2020, presents a set of KPIs for each of the axes addressed in the document. Among the covered domains, there is a strong orientation towards the use of AI to improve the population's well-being through the enhancement of public services, with the definition of various KPIs for this front. \\
    \hline
    South Africa & Does not have a strategy specifically defined for Artificial Intelligence. It is worth mentioning, however, that the country recently created the national artificial intelligence institute (Artificial Intelligence Institute of SA - AIISA), one of the initiatives planned for the development of the Fourth Industrial Revolution in the country. \\
    \hline
    South Korea & South Korea has stood out for significant government investment in AI, aiming to become a world leader in AI by 2030. In 2019, the country launched the ``Artificial Intelligence Development Strategy,'' \cite{KOR_AI_Strategy} which includes detailed plans to promote AI research, development, and application in various sectors such as health, transportation, manufacturing, and public services. A highlight of the South Korean strategy is the innovation in startups and national AI talent development, with public-private support and an emphasis on AI education and training. Additionally, the South Korean government has been seeking to boost AI-based economic growth, with detailed and action-oriented public policies, including performance rankings and public-private partnerships to promote innovation in the AI field. \\
    \hline
    United Kingdom & The British strategy (National AI Strategy) \cite{GBR_AI_Strategy} presents short, medium, and long-term lines of action but does not define indicators. The document mentions that the creation of indicators will be done in the future. \\
    \hline
    USA & The US strategy was published in 2016. The original document includes a set of broad guidelines for AI development. No set of monitoring indicators is defined. An annual report published in 2020 recaps some guidelines and mentions some goals for which indicators can be envisioned, although the report does not explicitly define them. Finally, under the Biden administration, the National Security Commission presents a long report in which new strategic elements are defined in the civil and defense realms. There is no explicit definition of indicators. In our investigation, we considered metrics from the three mentioned documents, even though such metrics were not explicitly defined as indicators. \\
    \hline
\end{longtable}
\normalsize

\begin{figure}[H]
    \centering
    \includegraphics[width=0.8\textwidth]{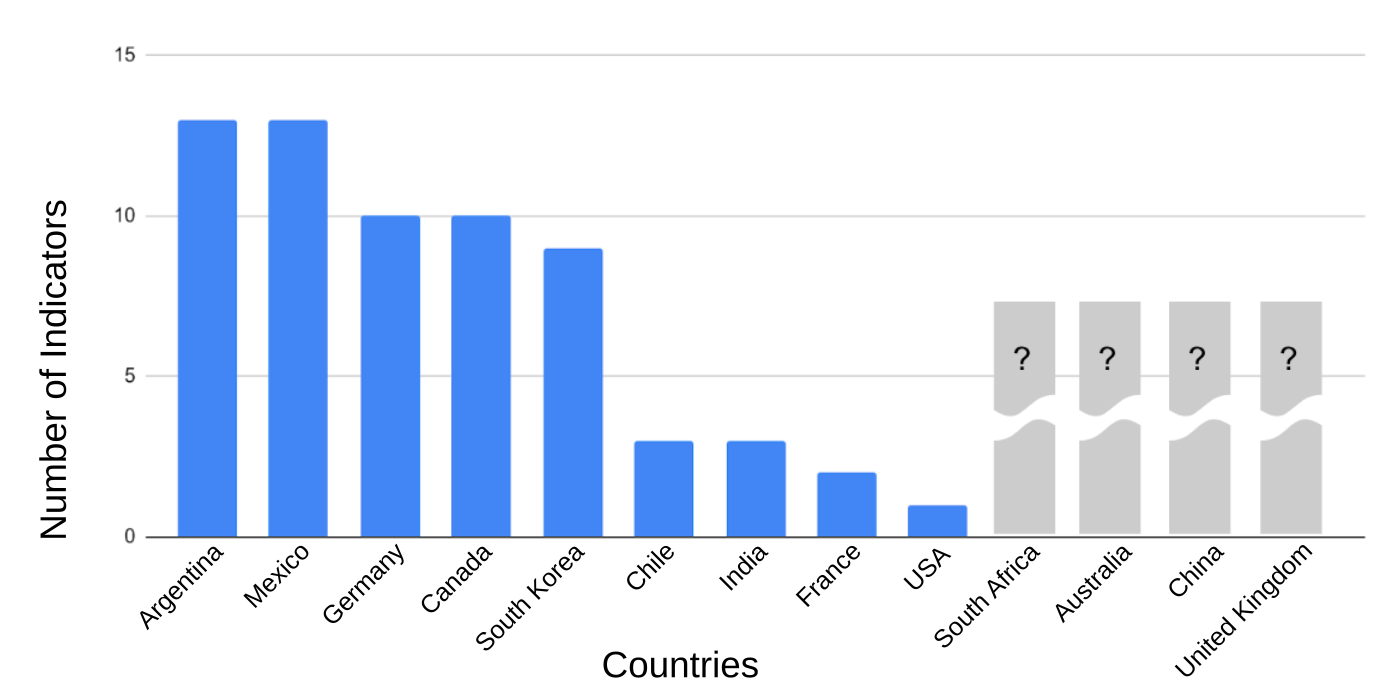}
    \caption{Number of proposed indicators found in the strategies of the analyzed countries.}
    \label{fig:nro_indicators}
\end{figure}

After analyzing the selected strategies in terms of the adoption of indicators (Activity 1.4), we obtained the results shown in Figure \ref{fig:nro_indicators}. This figure presents the absolute frequencies of the correspondences established between the proposed indicators and those present in the analyzed strategies. It was observed that strategies of 5 countries systematically rely on indicators - Germany, Argentina, Canada, South Korea, and Mexico. For a more detailed examination of the strategic use of indicators by reference countries in Stage 2, we can classify the sampled countries into the following strata:

\begin{itemize}
\item {NAIS that systematically rely on indicators:} Germany, Argentina, Canada, South Korea, Mexico.
\item {NAIS that do not rely on indicators but plan to do so:} France, India, United Kingdom.
\item {NAIS that neither rely on indicators nor plan to do so:} Australia, China, Chile, USA.
\item {Countries without NAIS:} South Africa.
\end{itemize}

It is worth noting that the standout strategies include representatives from four continents, with the two largest Latin American economies after Brazil, one of which is a member of Mercosur. This identifies a scope that is both robust and diverse for the subsequent analyses.

The next step in the proposed methodology is to perform a prevalence analysis of the indicators from the initial list in the analyzed strategies. The correspondence between the preliminary list of indicators and those found in the analyzed strategies is presented in Table \ref{tab_indicadoresdalistanasENIAs}. In this table, we associate each country with a hyperlink to its relevant document. If only a partial match is found between a proposed indicator and an indicator in the sample document, the similarity will be highlighted in yellow.

\begin{longtable}
{|>{\tiny}m{0.04\textwidth}|>{\tiny}m{0.04\textwidth}|>{\tiny}m{0.04\textwidth}|>{\tiny}m{0.04\textwidth}|>{\tiny}m{0.04\textwidth}|>{\tiny}m{0.04\textwidth}|>{\tiny}m{0.04\textwidth}|>{\tiny}m{0.04\textwidth}|>{\tiny}m{0.04\textwidth}|>{\tiny}m{0.04\textwidth}|>{\tiny}m{0.04\textwidth}|>{\tiny}m{0.04\textwidth}|>{\tiny}m{0.04\textwidth}|}
\caption{Prevalence of indicators from the initial list in the analysed NSAI. The page number refers to the analyzed documents for each strategy. The corresponding documents for each country are: DEU~\cite{DEU_AI_Strategy}, ARG~\cite{ARG_AI_Strategy}, AUS~\cite{AUS_AI_Strategy}, CAN~\cite{CAN_AI_Strategy}, CHL~\cite{CHL_AI_Strategy}, CHN~\cite{CHN_AI_Strategy}, KOR~\cite{KOR_AI_Strategy}, USA~\cite{USA_AI_Strategy}, FRA~\cite{FRA_AI_Strategy_Phase2}, IND~\cite{IND_AI_Strategy}, MEX~\cite{MEX_AI_Strategy} and GBR~\cite{GBR_AI_Strategy}.}\label{tab_indicadoresdalistanasENIAs}\\
\hline
    \textbf{KPI} 
    & \textbf{DEU}
    & \textbf{ARG} 
    & \textbf{AUS}
    & \textbf{CAN} 
    & \textbf{CHL}  
    & \textbf{CHN} 
    & \textbf{KOR} 
    & \textbf{USA}
    & \textbf{FRA}  
    & \textbf{IND}
    & \textbf{MEX}
    & \textbf{GBR} \\\hline
    A01 & & &&&&&&&&&&\\\hline
    A02 & & \cellcolor{yellow!25}\href{https://oecd-opsi.org/wp-content/uploads/2021/02/Argentina-National-AI-Strategy.pdf#page=71}{p.71} &&&&&&&&&&\\\hline
    A03 & &  &&&&&&&&&&\\\hline
    A04 & &  &&&&& \cellcolor{yellow!25}\href{https://wp.oecd.ai/app/uploads/2021/12/Korea_National_Strategy_for_Artificial_Intelligence_2019.pdf#page=18}{p.18} &&&&\cellcolor{yellow!25}\href{https://wp.oecd.ai/app/uploads/2022/01/Mexico_Agenda_Nacional_Mexicana_de_IA_2030.pdf#page=36}{p.36}; \href{https://wp.oecd.ai/app/uploads/2022/01/Mexico_Agenda_Nacional_Mexicana_de_IA_2030.pdf#page=90}{p.90}; \href{https://wp.oecd.ai/app/uploads/2022/01/Mexico_Agenda_Nacional_Mexicana_de_IA_2030.pdf#page=105}{p.105}&\\\hline
    A05 & & \cellcolor{yellow!25}\href{https://oecd-opsi.org/wp-content/uploads/2021/02/Argentina-National-AI-Strategy.pdf#page=143}{p.143} &&&&&  &&&&\cellcolor{yellow!25}\href{https://wp.oecd.ai/app/uploads/2022/01/Mexico_Agenda_Nacional_Mexicana_de_IA_2030.pdf#page=36}{p.36}; \href{https://wp.oecd.ai/app/uploads/2022/01/Mexico_Agenda_Nacional_Mexicana_de_IA_2030.pdf#page=90}{p.90}; \href{https://wp.oecd.ai/app/uploads/2022/01/Mexico_Agenda_Nacional_Mexicana_de_IA_2030.pdf#page=105}{p.105}&\\\hline
    A06 & &  &&&&&&&&&&\\\hline
    A07 & \cellcolor{yellow!25}{\href{https://www.bmwk.de/Redaktion/DE/Publikationen/Wirtschaft/einsatz-von-ki-deutsche-wirtschaft.pdf?__blob=publicationFile&v=1#page=5}{p.5}; \href{https://www.bmwk.de/Redaktion/DE/Publikationen/Wirtschaft/einsatz-von-ki-deutsche-wirtschaft.pdf?__blob=publicationFile&v=1#page=6}{p.6}; \href{https://www.bmwk.de/Redaktion/DE/Publikationen/Wirtschaft/einsatz-von-ki-deutsche-wirtschaft.pdf?}{site}} & \cellcolor{yellow!25}\href{https://oecd-opsi.org/wp-content/uploads/2021/02/Argentina-National-AI-Strategy.pdf#page=161}{p.161} &&&&&&&&&\href{https://wp.oecd.ai/app/uploads/2022/01/Mexico_Agenda_Nacional_Mexicana_de_IA_2030.pdf#page=90}{p.90}&\\\hline 
    A08 & \cellcolor{yellow!25}{\href{https://www.bmwk.de/Redaktion/DE/Publikationen/Wirtschaft/einsatz-von-ki-deutsche-wirtschaft.pdf?__blob=publicationFile&v=1#page=5}{p.5}; \href{https://www.bmwk.de/Redaktion/DE/Publikationen/Wirtschaft/einsatz-von-ki-deutsche-wirtschaft.pdf?__blob=publicationFile&v=1#page=15}{p.15}}&  &&&&&&&&&&\\\hline
    A09 & \cellcolor{yellow!25}{\href{https://www.bmwk.de/Redaktion/DE/Publikationen/Wirtschaft/einsatz-von-ki-deutsche-wirtschaft.pdf?__blob=publicationFile&v=1#page=6}{p.6}; \href{https://www.bmwk.de/Redaktion/DE/Publikationen/Wirtschaft/einsatz-von-ki-deutsche-wirtschaft.pdf?__blob=publicationFile&v=1#page=16}{p.16}; \href{https://www.bmwk.de/Redaktion/DE/Publikationen/Wirtschaft/einsatz-von-ki-deutsche-wirtschaft.pdf?__blob=publicationFile&v=1#page=17}{p.17}}&  &&&&&&&&&&\\\hline
    A10 & &  &&&&&&&&&&\\\hline
    A11 & &  &&&&&&&&&&\\\hline
    A12 & &  &&&&&&&&&&\\\hline
    B01 & & \cellcolor{yellow!25}\href{https://oecd-opsi.org/wp-content/uploads/2021/02/Argentina-National-AI-Strategy.pdf#page=118}{p.118} &&&&&&&\cellcolor{yellow!25}\href{https://www.aiforhumanity.fr/pdfs/9782111457089_Rapport_Villani_accessible.pdf#page=80}{p.80}&\cellcolor{yellow!25}\href{https://niti.gov.in/sites/default/files/2019-01/NationalStrategy-for-AI-Discussion-Paper.pdf#=55}{p.55}; \href{https://niti.gov.in/sites/default/files/2019-01/NationalStrategy-for-AI-Discussion-Paper.pdf#=75}{p.75} &\href{https://wp.oecd.ai/app/uploads/2022/01/Mexico_Agenda_Nacional_Mexicana_de_IA_2030.pdf#page=89}{p.89}; \href{https://wp.oecd.ai/app/uploads/2022/01/Mexico_Agenda_Nacional_Mexicana_de_IA_2030.pdf#page=90}{p.90}; \href{https://wp.oecd.ai/app/uploads/2022/01/Mexico_Agenda_Nacional_Mexicana_de_IA_2030.pdf#page=105}{p.105}&\\\hline
    B02 & &  &&\cellcolor{yellow!25}\href{https://cifar.ca/wp-content/uploads/2020/11/Pan-Canadian-AI-Strategy-Impact-Assessment-Report.pdf#page=9}{p.9}&&&&&&&&\\\hline
    B03 & &  &&&&&&&&&&\\\hline
    B04 & &  &&&&&&&&&&\\\hline
    B05 & &  &&\cellcolor{yellow!25}\href{https://cifar.ca/wp-content/uploads/2020/11/Pan-Canadian-AI-Strategy-Impact-Assessment-Report.pdf#page=9}{p.9}&&&&&&&&\\\hline
    B06 & \cellcolor{yellow!25}\href{https://www.plattform-lernende-systeme.de/ai-monitoring.html}{site} &  &&&&&&\href{https://www.nscai.gov/wp-content/uploads/2021/03/Full-Report-Digital-1.pdf#page=205}{p.205 (NSC strategy)}&&&\href{https://wp.oecd.ai/app/uploads/2022/01/Mexico_Agenda_Nacional_Mexicana_de_IA_2030.pdf#page=117}{p.117}&\\\hline
    B07 & &  &&&&&&&&&&\\\hline
    B08 & &  &&&&&&&&&&\\\hline
    B09 & &  &&&&&&&&&&\\\hline
    B10 & &  &&&&&&&&&&\\\hline
    B11 & &  &&&&&&&&&&\\\hline
    B12 & &  &&&&&&&&&&\\\hline
    B13 & & \href{https://oecd-opsi.org/wp-content/uploads/2021/02/Argentina-National-AI-Strategy.pdf#page=118}{p.118} &&&&&&&\cellcolor{yellow!25}\href{https://www.aiforhumanity.fr/pdfs/9782111457089_Rapport_Villani_accessible.pdf#page=120}{p.120}&&\cellcolor{yellow!25}\href{https://wp.oecd.ai/app/uploads/2022/01/Mexico_Agenda_Nacional_Mexicana_de_IA_2030.pdf#page=89}{p.89}&\\\hline
    B14 & &  &&&&&&&&&\cellcolor{yellow!25}\href{https://wp.oecd.ai/app/uploads/2022/01/Mexico_Agenda_Nacional_Mexicana_de_IA_2030.pdf#page=89}{p.89}&\\\hline
    B15 & &  &&&&&&&&&\cellcolor{yellow!25}\href{https://wp.oecd.ai/app/uploads/2022/01/Mexico_Agenda_Nacional_Mexicana_de_IA_2030.pdf#page=89}{p.89}&\\\hline
    B16 & &  &&&&&&&&\href{https://niti.gov.in/sites/default/files/2019-01/NationalStrategy-for-AI-Discussion-Paper.pdf#=51}{p.51}&&\\\hline
    B17 & & \href{https://oecd-opsi.org/wp-content/uploads/2021/02/Argentina-National-AI-Strategy.pdf#page=118}{p.118} &&\href{https://cifar.ca/wp-content/uploads/2020/11/Pan-Canadian-AI-Strategy-Impact-Assessment-Report.pdf#page=9}{p.9}&&&&&&\href{https://niti.gov.in/sites/default/files/2019-01/NationalStrategy-for-AI-Discussion-Paper.pdf#=51}{p.51}&&\\\hline
    B18 & &  &&\href{https://cifar.ca/wp-content/uploads/2020/11/Pan-Canadian-AI-Strategy-Impact-Assessment-Report.pdf#page=9}{p.9}&&&&&&&&\\\hline
    B19 & &  &&&&&&&&&&\\\hline
    B20 & \cellcolor{yellow!25}\href{https://www.plattform-lernende-systeme.de/ai-monitoring.html}{site}& \cellcolor{yellow!25}\href{https://oecd-opsi.org/wp-content/uploads/2021/02/Argentina-National-AI-Strategy.pdf#page=72}{p.72}&&&&&&&&&&\\\hline
    B21 & &  && \cellcolor{yellow!25}\href{https://cifar.ca/wp-content/uploads/2020/11/Pan-Canadian-AI-Strategy-Impact-Assessment-Report.pdf#page=12}{p.12}&&&&&&&\href{https://wp.oecd.ai/app/uploads/2022/01/Mexico_Agenda_Nacional_Mexicana_de_IA_2030.pdf#page=91}{p.91}&\\\hline
    B22 & \href{https://www.plattform-lernende-systeme.de/files/Downloads/Diverses/Studie_KI_in_Studium_und_Lehre.pdf#page=39}{p.39}; \href{https://www.plattform-lernende-systeme.de/ai-monitoring.html}{site}& \href{https://oecd-opsi.org/wp-content/uploads/2021/02/Argentina-National-AI-Strategy.pdf#page=71}{p.71}; \href{https://oecd-opsi.org/wp-content/uploads/2021/02/Argentina-National-AI-Strategy.pdf#page=118} {p.118} &&&&&&&&&&\\\hline
    B23 & & \href{https://oecd-opsi.org/wp-content/uploads/2021/02/Argentina-National-AI-Strategy.pdf#page=71}{p.71} &&& \href{https://www.minciencia.gob.cl/uploads/filer_public/bc/38/bc389daf-4514-4306-867c-760ae7686e2c/documento_politica_ia_digital_.pdf#page=40}{p.40}&&&&&&&\\\hline
    B24 & & \href{https://oecd-opsi.org/wp-content/uploads/2021/02/Argentina-National-AI-Strategy.pdf#page=71}{p.71} &&& \href{https://www.minciencia.gob.cl/uploads/filer_public/bc/38/bc389daf-4514-4306-867c-760ae7686e2c/documento_politica_ia_digital_.pdf#page=41}{p.41}&&&&&&&\\\hline
    B25 & &  &&&&&&&&&&\\\hline
    B26 & &  &&&&&&&&&&\\\hline
    B27 & \cellcolor{yellow!25}\href{https://www.plattform-lernende-systeme.de/ai-monitoring.html}{site}&  &&&&&&&&&&\\\hline
    B28 & &  &&&&&&&&&&\\\hline
    B29 & &  &&&&&&&&&&\\\hline
    B30 & &  &&&&&&&&&&\\\hline
    C01 & & \cellcolor{yellow!25}\href{https://oecd-opsi.org/wp-content/uploads/2021/02/Argentina-National-AI-Strategy.pdf#page=71}{p.71} && \cellcolor{yellow!25}\href{https://cifar.ca/wp-content/uploads/2020/11/Pan-Canadian-AI-Strategy-Impact-Assessment-Report.pdf#page=11}{p.11} && \href{https://wp.oecd.ai/app/uploads/2021/12/Korea_National_Strategy_for_Artificial_Intelligence_2019.pdf#page=31}{p.31}&&&&&\href{https://wp.oecd.ai/app/uploads/2022/01/Mexico_Agenda_Nacional_Mexicana_de_IA_2030.pdf#page=41}{p.41}&\\\hline
    C02 & \href{https://www.plattform-lernende-systeme.de/files/Downloads/Diverses/Studie_KI_in_Studium_und_Lehre.pdf#page=24}{p.24}; \href{https://www.plattform-lernende-systeme.de/ai-monitoring.html}{site}& \href{https://oecd-opsi.org/wp-content/uploads/2021/02/Argentina-National-AI-Strategy.pdf#page=71}{p.71} && \cellcolor{yellow!25}\href{https://cifar.ca/wp-content/uploads/2020/11/Pan-Canadian-AI-Strategy-Impact-Assessment-Report.pdf#page=12}{p.12}&& \cellcolor{yellow!25}\href{https://wp.oecd.ai/app/uploads/2021/12/Korea_National_Strategy_for_Artificial_Intelligence_2019.pdf#page=31}{p.31}&&&&&\cellcolor{yellow!25}\href{https://wp.oecd.ai/app/uploads/2022/01/Mexico_Agenda_Nacional_Mexicana_de_IA_2030.pdf#page=89}{p.89}&\\\hline
    C03 & \cellcolor{yellow!25}\href{https://www.plattform-lernende-systeme.de/ai-monitoring.html}{site}& \cellcolor{yellow!25}\href{https://oecd-opsi.org/wp-content/uploads/2021/02/Argentina-National-AI-Strategy.pdf#page=71}{p.71} &&&& \href{https://wp.oecd.ai/app/uploads/2021/12/Korea_National_Strategy_for_Artificial_Intelligence_2019.pdf#page=31}{p.31} &&&&&&\\\hline
    C04 & &  &&&&&&&&&&\\\hline
    C05 & &  &&&&&&&&&&\\\hline
    C06 & &  &&&& \cellcolor{yellow!25}\href{https://wp.oecd.ai/app/uploads/2021/12/Korea_National_Strategy_for_Artificial_Intelligence_2019.pdf#page=33}{p.33}&&&&&&\\\hline
    C07 & &  &&&&&&&&&&\\\hline
    C08 & &  &&&&&&&&&&\\\hline
    D01 & \cellcolor{yellow!25}\href{https://www.bmwk.de/Redaktion/DE/Publikationen/Wirtschaft/einsatz-von-ki-deutsche-wirtschaft.pdf?__blob=publicationFile&v=1#page=5}{p.5}; \href{https://www.bmwk.de/Redaktion/DE/Publikationen/Wirtschaft/einsatz-von-ki-deutsche-wirtschaft.pdf?__blob=publicationFile&v=1#page=33}{p.33}; \href{https://www.bmwk.de/Redaktion/DE/Publikationen/Wirtschaft/einsatz-von-ki-deutsche-wirtschaft.pdf?__blob=publicationFile&v=1#page=34}{p.34}; &  && \href{https://cifar.ca/wp-content/uploads/2020/11/Pan-Canadian-AI-Strategy-Impact-Assessment-Report.pdf#page=10}{p.10}&& \href{https://wp.oecd.ai/app/uploads/2021/12/Korea_National_Strategy_for_Artificial_Intelligence_2019.pdf#page=30}{p.30}&&&&&\cellcolor{yellow!25}\href{https://wp.oecd.ai/app/uploads/2022/01/Mexico_Agenda_Nacional_Mexicana_de_IA_2030.pdf#page=89}{p.89}&\\\hline
    D02 & &  && \href{https://cifar.ca/wp-content/uploads/2020/11/Pan-Canadian-AI-Strategy-Impact-Assessment-Report.pdf#page=10}{p.10}&&&&&&&\cellcolor{yellow!25}\href{https://wp.oecd.ai/app/uploads/2022/01/Mexico_Agenda_Nacional_Mexicana_de_IA_2030.pdf#page=89}{p.89}&\\\hline
    E01 & &  &&& \href{https://www.minciencia.gob.cl/uploads/filer_public/bc/38/bc389daf-4514-4306-867c-760ae7686e2c/documento_politica_ia_digital_.pdf#page=41}{p.41}&&&&&&&\\\hline
    E02 & &  &&&& \href{https://wp.oecd.ai/app/uploads/2021/12/Korea_National_Strategy_for_Artificial_Intelligence_2019.pdf#page=12}{p.12}&&&&&&\\\hline
    F01 & &  && \href{https://cifar.ca/wp-content/uploads/2020/11/Pan-Canadian-AI-Strategy-Impact-Assessment-Report.pdf#page=14}{p.14}&& \href{https://wp.oecd.ai/app/uploads/2021/12/Korea_National_Strategy_for_Artificial_Intelligence_2019.pdf#page=14}{p.14}&&&&&&\\\hline
    F02 & &  &&&& \href{https://wp.oecd.ai/app/uploads/2021/12/Korea_National_Strategy_for_Artificial_Intelligence_2019.pdf#page=10}{p.10}&&&&&&\\\hline
\end{longtable}

\begin{figure}[H]
    \centering
    \includegraphics[width=0.8\textwidth]{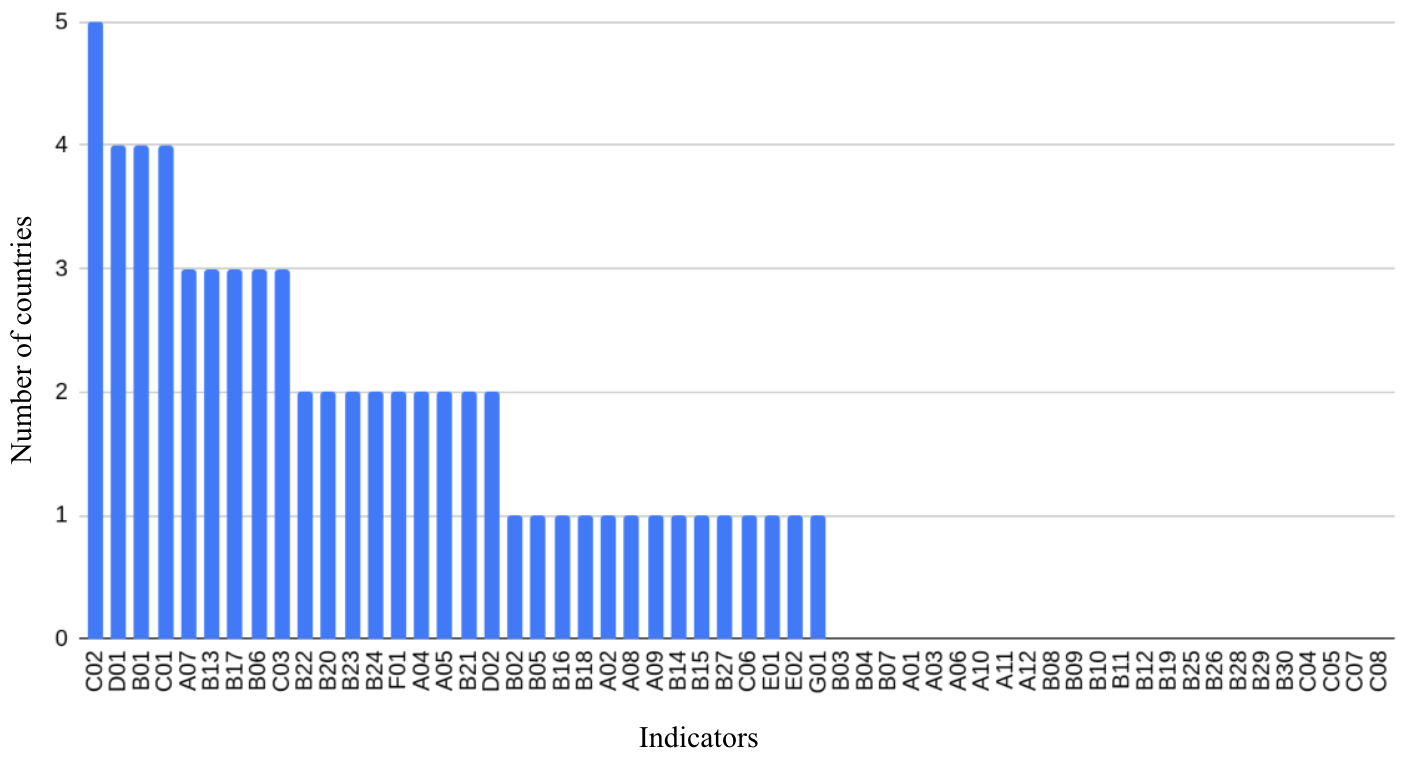}
    \caption{Number of countries using each of the proposed indicators.}\label{Fig:freq_indicador_pais}
\end{figure}

Figure \ref{Fig:freq_indicador_pais} shows the number of countries that adopt each of the indicators defined by CGEE, revealing that most of the indicators are not considered, and only a fraction is used by three or more countries. This sparsity in Table \ref{tab_indicadoresdalistanasENIAs} can be attributed to two factors, which we will now discuss.

It is notable, for instance, that indicator D01 (AI skills required by companies in job vacancies) is among the most frequently used, appearing in five countries out of the total analyzed. However, only nine countries’ strategies rely on indicators or foresee the future definition of indicators (as shown in Figure \ref{fig:nro_indicators}). Among those countries planning to define indicators, it is not always clear what data sources will be used, which complicates the establishment of precise correspondences.

The sparse nature of Table \ref{tab_indicadoresdalistanasENIAs} also stems, to some extent, from the high degree of specificity in the indicators initially proposed by CGEE. Even in strategies that clearly define their indicators, these are often developed without extensive refinement of quantitative data. Despite this sparsity, we can still draw meaningful conclusions about the connections between the analyzed strategies and the proposed indicators, particularly when considering their aggregation at the “Dimension” level, as will be discussed.

As a result of Activity 1.5, Table \ref{tab_prevalence} relates the prevalence of the identified indicators to the seven analytical dimensions (A-G) defined by CGEE, providing a qualitative synthesis for each case. Additionally, Table \ref{tab:highly-relevant-indicators} highlights the most relevant indicators.

\begin{table}[!ht]
    \centering\footnotesize
    \caption{Prevalence of the indicators found, among the seven analytical dimensions listed by CGEE (A-G).}\label{tab_prevalence}
    \begin{tabular}{|p{1.5cm}|p{1.4cm}|p{1.4cm}|p{1.4cm}|p{4cm}|}\hline
     \multicolumn{1}{|c|}{\multirow{2}{*}{\bf Dimension}}    & \multicolumn{3}{c|}{\bf Indicators, by frequency} & \multicolumn{1}{|c|}{\multirow{2}{*}{\bf Synthesis}} \\ 
                  & Irrelevant & Prevalent & Highly Prevalent & \\\hline
      Adoption and Use of AI Applications & A01, A03, A06, A10, A11, A12 & A02, A04, A05, A08, A09 & A07 & As a rule, specific indicators related to AI application areas were not identified, which explains the presence of indicators A01 (agriculture), A11 (health), and A12 (public security) in the "irrelevant" group. Additionally, while several analyzed strategies address education, there are few indicators corresponding to those defined in the initial CGEE study. On the other hand, there is some alignment between the initially defined indicators and those analyzed in strategies regarding AI adoption in public administration and businesses.\\\hline
      Knowledge Production & B03, B04, B07, B08, B09, B10, B11, B12, B19, B25, B26, B28, B29, B30 & B02, B05,  B14, B15, B16, B18, B20, B21,  B22, B23, B24, B27 & B01, B06, B13, B17 & Although some indicators have been classified as irrelevant, it is worth noting that these indicators are highly specific, such as B12 (Journals where AI researchers publish) and B26 (Publications in AI by AI topic). Therefore, we believe that, overall, the indicators in this dimension are relatively common in the investigated strategies, except for minor variations in definition.\\\hline
    \end{tabular}
\end{table}

\begin{table}[!ht]
\centering
\caption{Highly relevant indicators.}
\label{tab:highly-relevant-indicators}
\begin{tabular}{|l|l|}
\hline
\textbf{Acronym} & \textbf{Indicator} \\ \hline
A07 & Companies Using Artificial Intelligence Technologies \\ \hline
B01 & Networks of Scientific and Academic Collaboration \\ \hline
B06 & Patents in AI - Total \\ \hline
B13 & Total Number of Researchers in AI \\ \hline
B17 & Articles and Papers on AI \\ \hline
C01 & Number of Postgraduates in AI-related Areas \\ \hline
C02 & Number of Postgraduate Programs in AI-related Areas \\ \hline
C03 & Technical Programming Courses \\ \hline
D01 & AI Skills Required by Companies in Job Vacancies \\ \hline
\end{tabular}
\end{table}

In the next step (Activity 1.6), we identified indicators present in the analyzed strategies that were not included in the preliminary list provided by CGEE. These indicators were selected based on their impact and particular relevance within the observed NAIS. For reference accuracy, we initially present them in Table \ref{tab:indicators_by_country} in the original language of each document.

\begin{table}
\centering \footnotesize
\caption{List of indicators identified in the analyzed NAIS which are not found in the preliminary list of CGEE.}
\label{tab:indicators_by_country}
\begin{tabular}{|p{0.15\textwidth}|p{0.7\textwidth}|}
\hline
\textbf{Country} & \textbf{Indicators} \\ \hline
Germany & 
Funding by the Federal Government\newline
Grant Programs to promote AI in SMEs\newline
AI Startups\newline
Publicly funded Transfer Hubs\\ \hline
Argentina & 
     Inversión total en Actividades de Innovación\newline
     Potencia de cálculo disponible en el país\newline
     Cantidad de organizaciones con infraestructura de supercómputo orientada a actividades de formación, I+D+i, implementación proyectos IA\newline
     Capacidad de supercómputo en instituciones públicas\newline
     Capacidad de organizaciones privadas con capacidad de supercómputo orientada a IA\newline
     Cantidad de laboratorios y grupos de investigación interdisciplinarios de IA\newline
     Cantidad de nuevas carreras específicas IA\newline
     Cantidad de Instituciones no formales con formación a temáticas asociadas a IA\newline
     Cantidad Institutos de Investigación orientados a temáticas asociadas a IA\newline
     Cantidad de Becarios con especialización en temáticas asociadas a IA\newline
     Cantidad de sitios o plataformas seguras para poder compartir bases de datos con el fin de ser utilizadas para fomentar el desarrollo de sistemas basados en IA\\ \hline
Canada & 
     AI Startup Funding\newline
     National Technology spend on AI\newline
     \# AI companies\newline
     Government investment in IA\newline
     AI Skills Migration Index\newline
     Public Sentiment of AI \\ \hline
South Korea & 
     AI technology competitiveness\newline
     Economic effect of AI (McKinsey)\newline
     Number of open data cases\newline
     AI startup funding\\ \hline
Mexico & 
     Cantidad de cursos y certificaciones en IA\newline
     Número de becas relacionadas con la IA\newline
     Número de personas pertenecientes a minorías involucradas en IA\newline
     Monto total y distribución de recursos asignados a I+D en IA (público y privado)\newline
     Número de centros de investigación, instituciones, empresas y 
     dependencias gubernamentales donde se desarrolla o usa IA\newline
     Número de iniciativas o protector de IA sociales \newline
     Número de expertos repatriados o atraídos \\ \hline
\end{tabular}
\end{table}

Then, the identified indicators were categorized, whenever possible, according to the taxonomy of indicators adopted in the preliminary list; for genuinely singular cases, we proposed new hierarchies.

One analytical perspective observed in the highlighted sample strategies, which was absent from the preliminary study by CGEE, concerns the financial aspect, including investments in AI or sector revenues, both from public and private actors. The new indicators that measure these aspects are A14, A16, I01, I02, and I03. Financial and infrastructural aspects hold significant prominence in the observed national strategies, justifying the introduction of two new dimensions to the taxonomy initially proposed by CGEE: “Investments in R\&DI” and “Centers, hubs, and multi-user structures.” The identified indicators and their dimensions are presented below.

\paragraph{Existing Dimensions}
\begin{itemize}
    \item \textbf{Adoption and use of AI-based applications}
    \begin{itemize}
        \item Number of startups operating in AI
        \item Investment in SMEs and startups operating in AI - total
        \item Number of companies with R\&D departments working on AI
        \item Revenue from AI-based products and services (by type)
        \item Knowledge production
        \item Young or emerging research groups in AI
    \end{itemize}
    \item \textbf{Training}
    \begin{itemize}
        \item Number of AI scholarships funded by the private sector
        \item Number of AI-oriented training centers
    \end{itemize}
    \item \textbf{Skills and Employment}
    \begin{itemize}
        \item Number of AI experts repatriated or recruited
        \item Monitoring of international agendas
        \item AI technology competitiveness
        \item Monitoring of political debates and regulatory aspects
        \item Proportion of women or minority groups working in AI
        \item Number of social projects
        \item Public sentiment scale regarding AI development
        \item Ethical pillars in AI
    \end{itemize}
\end{itemize}

\paragraph{New Dimensions}
\begin{itemize}
    \item \textbf{Centers, Hubs, and Multi-user Structures}
    \begin{itemize}
        \item Number of large high-performance computing centers for AI
        \item Technology transfer hubs
        \item Number of centers of excellence in AI-oriented research
        \item Number of secure data sharing portals or platforms oriented toward the development of AI-based systems
        \item Number of open data cases
    \end{itemize}
    \item \textbf{Investment in R\&DI}
    \begin{itemize}
        \item Public investment in AI R\&D
        \item Private investment in AI R\&D
        \item Investment in SMEs or startups operating in AI - innovation promotion and subsidy programs
    \end{itemize}
\end{itemize}

Furthermore, new areas within the existing dimensions were also identified during this Stage, particularly those addressing factors related to the maturity of the AI innovation ecosystem and perspectives on Responsible AI concepts and public opinion.

The outcome of Stage 1 is a consolidated set of feasible indicators, which includes both those initially proposed in the preliminary list and those highly prevalent across the observed NAIS, as well as new indicators identified from standout NAIS. The final list of relevant strategic AI indicators is presented in Table \ref{tab:consolidated-indicators}.
\begin{table}[!ht]
    \centering \footnotesize
    \caption{Consolidated set of feasible indicators.}\label{tab:consolidated-indicators}
    \begin{tabular}{|c|p{0.8\textwidth}|}
        \hline
        \textbf{Acronym} & \textbf{Indicator} \\
        \hline
        A07 & Companies Using Artificial Intelligence Technologies \\
        \hline
        A13 & Number of startups operating in AI \\
        \hline
        A14 & Investment in SMEs and startups operating in AI - total \\
        \hline
        A15 & Number of companies with R\&D departments working on AI \\
        \hline
        A16 & Revenue from AI-based products and services (by type) \\
        \hline
        B01 & Scientific and academic collaboration networks \\
        \hline
        B06 & AI patents - total \\
        \hline
        B13 & Total number of AI researchers \\
        \hline
        B17 & AI-related articles and papers \\
        \hline
        B31 & Young or emerging AI research groups \\
        \hline
        C01 & Number of postgraduates in AI-related fields \\
        \hline
        C02 & Number of postgraduate programs in AI-related fields \\
        \hline
        C03 & Technical programming courses \\
        \hline
        C09 & Number of AI scholarships funded by the private sector \\
        \hline
        C10 & Number of AI-oriented training centers \\
        \hline
        D01 & AI skills required by companies in job openings \\
        \hline
        D03 & Number of AI experts repatriated or recruited \\
        \hline
        E03 & AI technology competitiveness \\
        \hline
        F02 & Proportion of women or minority groups working in AI \\
        \hline
        F03 & Number of social projects \\
        \hline
        F04 & Public sentiment scale regarding AI development \\
        \hline
        F05 & Ethical pillars in AI \\
        \hline
        H01 & Number of large high-performance computing centers for AI \\
        \hline
        H02 & Technology transfer hubs \\
        \hline
        H03 & Number of centers of excellence in AI-oriented research \\
        \hline
        H04 & Number of secure data sharing portals or platforms oriented toward the development of AI-based systems \\
        \hline
        H05 & Number of open data cases \\
        \hline
        I01 & Public investment in AI R\&D \\
        \hline
        I02 & Private investment in AI R\&D \\
        \hline
        I03 & Investment in SMEs and startups operating in AI - innovation promotion and subsidy programs \\
        \hline
    \end{tabular}
\end{table}

\subsection{Stage 2}\label{subsec:Stage2br}

In Stage 2, we aimed to align the relevant indicators with the structure of the analyzed strategy. Initially, we considered quantifying this by counting the number of strategic actions linked to each indicator within each EBIA axis. However, the number of strategic actions per axis varies significantly (from 5 to 15), meaning that the number of correlations to be checked against 30 indicators across 9 axes would range from 1,350 to 4,050. Since this process requires manual conceptual analysis and cannot be automated, we determined it would be impractical. Additionally, the usefulness of such detailed counting for public policy remains unclear.

Therefore, as a first approximation, we adopt a simple binary correspondence analysis between indicators and axes, considering them related whenever an indicator directly serves the conceptual purpose of an axis or at least one of its specific strategic actions (defined in Table \ref{tab:actionsEBIA}). In particular, the conceptual purpose was formulated based on examination and summary of the corresponding detailed description in the EBIA text.
We also developed a resource for visualizing this correspondence. Given the dual nature of EBIA’s thematic structure -- between its ideal $3\times 6$ matrix arrangement and the set of 9 independent axes -- we adopted an extended $4\times 7$ matrix arrangement. This matrix situates each indicator at the intersection of a vertical and a transversal axis, where possible. When such intersections are not epistemologically feasible for a certain indicator, it is placed in an additional column or row, accounting for cases omitted by the original EBIA matrix structure.

Table \ref{tab:strategic-actions} maps the 30 indicators from the consolidated set (Table \ref{tab:consolidated-indicators}) against the vertical and transversal axes of EBIA, as depicted in Figure \ref{Fig:Eixos_EBIA}. Unaddressed cases are placed in the column “Outside of vertical axes” or in the row “Outside of transversal axes.” This extended $4\times 7$ matrix arrangement allows a comprehensive visualization of the indicator set in relation to the possibilities and limitations of intersectionality posed by the ideal matrix structure of EBIA.

\begin{table}[!ht]
    \centering\footnotesize
       \caption{Indicators identified as relevant vs EBIA's Axes.}
    \label{tab:strategic-actions}
    \begin{tabular}{|p{1cm}|p{1.0cm}|p{1cm}|p{1cm}|p{1cm}|p{1cm}|p{1cm}|p{1cm}|}
        \hline
        \textbf{Axis} & \textbf{EDU} & \textbf{W\&T} & \textbf{R\&DE} & \textbf{App.PS} & \textbf{App.PA} & \textbf{PS} & \textbf{OVA}\\
        \hline
        \textbf{LR\&E}& & &
        A15, F05 & 
        A13, A14 & 
        F05 & & 
        A07  \\
        \hline
        \textbf{GOV} & & F04& & &
        H04, H05 & &
        F03  
        \\
        \hline
        \textbf{INT} & B01 & &
        B17, E03 & & & &\\\hline
        \textbf{OTA} &
        B13, C01, C02, C10, D01 & 
        B13, C03, C09, D01, D03, F02 & 
        A07, B06, H02, H03, I01 & 
        E03, H02, I02, I03 & & 
        F05 & 
        A16, B31, H01 \\
        \hline
    \end{tabular}
 
\end{table}

\begin{table}[!ht]
    \centering\footnotesize
    \caption{Frequency of relevant indicators vs EBIA's Axes.}
    \label{tab:strategic-actions-2}
    \begin{tabular}{|p{1cm}|p{0.7cm}|p{0.7cm}|p{0.7cm}|p{1cm}|p{1cm}|p{0.7cm}|p{0.7cm}||p{0.7cm}|}
        \hline
        \textbf{Axis} & \textbf{EDU} & \textbf{W\&T} & \textbf{R\&DE} & \textbf{App.PS} & \textbf{App.PA} & \textbf{PS} & \textbf{OVA}&\textbf{\#Ind.}\\
        \hline
        \textbf{LR\&E}& & &
        2 & 
        2 & 
        1 & & 
        1 & 5 \\
        \hline
        \textbf{GOV} & & 1& & &
        2 & &
        1  & 4
        \\
        \hline
        \textbf{INT} & 1 & &
        2 & & & & & 3\\\hline
        \textbf{OTA} &
        5 & 
        6 & 
        5 & 
        4 & & 
        1 & 
        3 & 22\\
        \hline\hline
        \textbf{\#Ind.} & 6 & 7 & 9 & 6 & 3& 1 & 5 & \\\hline
    \end{tabular}
\end{table}

Table \ref{tab:strategic-actions-2} presents the frequency analysis associated with Table \ref{tab:strategic-actions}, highlighting the aggregate incidence of the indicator set across each vertical and transverse axis. As some indicators appear more than once within the same transverse axis, the values in the column ``\# Indicators per transverse axis'' do not necessarily correspond to the sums of each row in the table.

\subsection{Stage 3}\label{subsec:Stage3br}

In Stage 3, we highlight some patterns identified in the correspondence analysis of the previous stage.

At first glance, each EBIA axis is covered by some indicator; indeed, there are at least 3 indicators per axis, except for the ``Application in Public Sector,'' which has only one associated indicator. To this extent, the consolidated set of indicators encompasses all of EBIA's themes. On the other hand, the vast majority of indicator occurrences are outside the intersectionalities foreseen by EBIA's ideal $3 \times 6$ matrix arrangement. This suggests that EBIA's conceptual framework does not sufficiently accommodate the diversity of prevalent indicators among the observed NAIS.
Across all vertical axes, except ``Application in Public Sector,'' occurrences of indicators outside the ideal $3 \times 6$ matrix arrangement exceed the total occurrences within it, totalling a ratio of 22:12. This indicates that EBIA's composition of transverse axes does not significantly encompass the diversity of the consolidated set of indicators in almost all thematic verticals.

Conversely, the opposite phenomenon is observed without exception across transverse axes, totalling a ratio of 5:32. Furthermore, each transverse axis only exceeds the total of the corresponding interior of the $3 \times 6$ arrangement by at most one additional indicator. This suggests that EBIA's vertical axes are largely aligned with the international thematic perspective of indicator monitoring.
Additionally, three indicators are observed in the lower-right cell, completely outside the ideal $3 \times 6$ matrix arrangement of EBIA. This means that 10\% of the consolidated set of 30 indicators occupy a complete conceptual blind spot in relation to EBIA. They are:

\begin{itemize}
    \item A16: Revenue from AI-based products and services (by type)
    \item B31: Young or emerging AI research groups
    \item H01: Number of large high-performance computing centers for AI
\end{itemize}

\subsection{Elements for improving EBIA}\label{subsec:ImproveEBIA}

The benchmarking conducted yielded important findings that go beyond the fundamental objective of identifying relevant indicators on the international stage. Indeed, through detailed analysis of the NAIS, points were identified that could support the improvement of EBIA through a revision process capable of aligning it more closely with international best practices, as well as addressing specific issues within the Brazilian context.

A first point to highlight is the observation of different structures in the studied NAIS. Some are guided, from their conception, by clear and measurable quantitative indicators; others are organized qualitatively around national thematic interests. EBIA, by contrast, adopts a different structure based on general thematic axes, which are supposed to communicate through a matrix arrangement, unfolding into a set of approximately 50 strategic actions. We also note, in passing, that unlike some NAIS, EBIA does not initially assign its strategic actions to particular execution actors.

Due to its broader axes, EBIA's structure allowed for a good accommodation of the indicators identified in international benchmarking. However, it should be noted that this result would occur even for other sets of indicators analyzed in the study but not selected for Product 3. This highlights a potentially limiting characteristic of EBIA, namely the lack of more clearly defined focuses capable of guiding decision-making instances in the planning, execution, and monitoring of the strategy.

The matrix structure of EBIA, which relates the six vertical thematic axes and three transverse axes, is conceptually interesting as it allows for addressing complex aspects from a multi-sectoral perspective. Nevertheless, the conceptual detailing and strategic actions associated with each of the nine axes are presented independently, which does not adequately emphasize the expected intersections within a matrix arrangement.

Given these observations, some insights can be envisioned for a revision process of EBIA:
\begin{enumerate}
    \item More detailed and assertive definition of objectives, goals, actions, and monitoring strategies for EBIA. Building upon the conceptual framework of EBIA, there is room for delineating and deepening strategic actions, relying precisely on the indicators mentioned here to quantify the diagnosis and goals for public policies derived from the strategy.
    \item Implementation of an organization into axes that truly follows a matrix rationality in all aspects. The challenges of implementing an artificial intelligence strategy in Brazil, which presents a high level of complexity across various socioeconomic and regional perspectives, require in-depth dialogue between different levels of government, as well as close engagement with academia, the public and private innovation ecosystem, and the broader productive sector. The formulation of a truly matrix structure is an interesting tool to leverage synergies among different stakeholders involved in AI development, as well as to provide a mediation space between potentially conflicting issues arising from technological AI development.
\end{enumerate}

It should be emphasized that the full development of improvement point 2 requires dialogue with stakeholders. Building on existing working groups in the management of EBIA, a recurring schedule of sectoral forums can be envisioned to bring together different segments of society, possibly culminating in the periodic organization of a National Artificial Intelligence Conference.

\section{Conclusion}
\label{sec:Conclusion}

This study proposed a generic methodology for monitoring National Artificial Intelligence Strategies (NAIS), demonstrating its applicability through the Brazilian Artificial Intelligence Strategy (EBIA). The developed approach was structured into two main components: identifying relevant indicators in the international landscape and analyzing the alignment between these indicators and the strategic actions outlined in the evaluated strategy. This analysis not only assessed the effectiveness of the monitoring measures but also provided insights into the structural quality of the strategy. 

The application of the proposed methodology requires a set of preliminary information to support some of its stages. While this requirement can be seen as the main limitation of our proposal, information on NAIS, as discussed throughout this work, is now accessible (or indexed) through various sources, including the OECD AI Policy Observatory. Moreover, information regarding the local context of certain countries has also been made available by national observatories and local agencies.

Regarding the case study carried out in this work, the application to EBIA identified specific points that could support its improvement, both in terms of alignment with international best practices and in addressing issues specific to the Brazilian context. These results underscore the importance of periodic reviews of national strategies. The proposed methodology has a generic nature, making it adaptable and applicable to other NAIS, presenting itself as a valuable tool for governments seeking to monitor, evaluate, and strengthen their artificial intelligence strategies.

\theendnotes
\newpage

\section*{Appendix}

\footnotesize
\begin{longtable}{|p{0.05\textwidth}|p{0.2\textwidth}|p{0.6\textwidth}|}
    \caption{List of preliminary indicators provided by CGEE. (Dimension A: Adoption and use of AI-based applications; Dimension B: Knowledge production; Dimension C: Education and Training; Dimension D: Skills and Employment; Dimension E: Monitoring of International Agendas; Dimension F: Monitoring of Political Debates and Regulatory Aspects; Dimension G: Monitoring of Trends and Innovation)} \label{tab_lista_preliminar}\\
    \hline
    \textbf{ID} & \textbf{Area}  & \textbf{Indicator}  \\
    \hline
    \endfirsthead
    \caption[]{(Continued)} \\
    \hline
    \textbf{ID} & \textbf{Area}  & \textbf{Indicator}  \\
    \hline
    \endhead
    \hline
    \endfoot
    \hline
    \endlastfoot
    A1 & Agriculture & Adoption and use of AI applications \\
    \hline
    A2 & Education &  Basic education schools using learning environment or platform \\
    \hline
    A3 & Education &  Basic education schools by resources provided by virtual learning environment or platform \\
    \hline
    A4 & Government\footnote{Executive, Legislative, Judiciary} &  Federal and state public agencies, by use of AI technologies in the last 12 months\\
    \hline
    A5 & Government & Federal and state public agencies that used AI technologies in the last 12 months, by type  \\
    \hline
    A6 & Government &  Federal and state public agencies, by reasons for not using AI technologies In The last 12 months  \\
    \hline
    A7 & Industry, Trade and Services &  Companies that used AI technologies \\
    \hline
    A8 & Industry, Trade and Services &  Companies that used AI technologies, by type \\
    \hline
    A9 & Industry, Trade and Services & Companies that used AI technologies, by type of application  \\
    \hline
    A10 & Industry, Trade and Services & Companies that did not use AI technologies, by type of obstacle  \\
    \hline
    A11 & Medicine and Health & Health establishments, by type of technology used  \\
    \hline
    A12 & Public safety &  \\
    \hline
    B1 & Research Groups and Institutions & Scientific and Academic Collaboration Networks \\
    \hline
    B2 & Patents & Number of 'AI' Patents Granted, By Inventor  \\
    \hline
    B3 & Patents & Average Time Between Filing an 'AI' Patent and Its Grant by the Patent Office  \\
    \hline
    B4 & Patents & Number of 'AI' Patents Granted to Applicants \\
    \hline
    B5 & Patents & Characterization of AI Patents by Country of Origin \\
    \hline
    B6 & Patents & AI Companies' Patent Deposits - Total  \\
    \hline
    B7 & Patents & Share of AI Companies' Patent Deposits in Total Companies' Patent Deposits \\
    \hline
    B8 & Patents & AI companies' patent deposits by theme (field of study) \\
    \hline
    B9 & Patents & AI patents per 100,000 staff employed \\
    \hline
    B10 & Patents &  Number of AI depositors by CNAE \\
    \hline
    B11 & Patents & Number of AI patents by CNAE  \\
    \hline
    B12 & Journals & Journals where AI researchers publish  \\
    \hline
    B13 & Researchers &  Total number of AI researchers  \\
    \hline
    B14 & Researchers &  Number of AI researchers by area of knowledge  \\
    \hline
    B15 & Researchers & Number of AI researchers at each level of qualification  \\
    \hline
    B16 & Scientific Publications &  Number of scientific publications on AI involving Brazilian researchers  \\
    \hline
    B17 & Scientific Publications & Articles and papers on AI \\
    \hline
    B18 & Scientific Publications  & Citations in articles and papers on AI \\
    \hline
    B19 & Scientific Publications  & Co-citation networks and bibliographic tracking \\
    \hline
    B20 & Scientific Publications & Number of citations received by academic AI  \\
    \hline
    B21 & Scientific Publications & Analysis of the situation, evolution, and emergence of researched AI themes in Brazil (Lattes) and worldwide (WOS)  \\
    \hline
    B22 & Scientific Publications & Publications in AI - total  \\
    \hline
    B23 & Scientific Publications & Publications in AI in journals \\
    \hline
    B24 & Scientific Publications & AI publications at conferences  \\
    \hline
    B25 & Scientific Publications & Publications in AI by area of knowledge  \\
    \hline
    B26 & Scientific Publications & Publications in AI by AI theme (field of study)  \\
    \hline
    B27 & Researcher Networks & Co-authorship networks \\
    \hline
    B28 & Researcher Networks & Number of publications per collaboration (Institution or Country) \\
    \hline
    B29 & Researcher Networks & Number of publications per co-authorship (Researchers)  \\
    \hline
    B30 & Software Registration & Software registration \\
    \hline
    C1 & Undergraduate and Graduate Courses & Number of postgraduates in AI-related areas \\
    \hline
    C2 & Undergraduate and Graduate Courses & Number of postgraduate programs in AI-related areas  \\
    \hline
    C3 & Technical Programming Courses &  \\
    \hline
    C4 & Basic Education &  Internet user basic education teachers, by request for students to use digital technologies in educational activities in the Last 12 months - mathematical and scientific resources \\
    \hline
    C5 & Basic Education & Internet user basic education teachers, by request for students to use digital technologies in educational activities in the last 12 months - programming resources  \\
    \hline
    C6 & Basic Education &  Basic education teachers, by themes of activities conducted with students on safe, responsible, and critical use of the internet in the last 12 months  \\
    \hline
    C7 & Institutions and Enrollments &  \\
    \hline
    C8 & Cooperation Links (School-Enterprise) &  \\
    \hline
    D1 & Labor Demand & AI Skills Required by Companies in Job Vacancies \\
    \hline
    D2 & Labor Supply &  AI Skills Declared by Individuals on Platforms  \\
    \hline
    E1 & International Organizations' Documents & International documents \\
    \hline
    E2 & International Agenda Monitoring & National AI strategies in leading countries   \\
    \hline
    F1 & Legal Frameworks, Bills, Infra-legal Regulation & National Documents \\
    \hline
    G1 & Technological and Market Trends & Articles and Prospecting of Topics \\
    \hline
\end{longtable}

\newpage
\begin{longtable}{|>{\raggedright\arraybackslash}p{2.5cm}|p{3cm}|p{4.5cm}|}
\caption{Axis and strategic actions of EBIA.}\label{tab:actionsEBIA} \\\hline
    \textbf{Axis} & \textbf{Conceptual Purpose} & \textbf{Strategic Actions} \\
    \hline
        LEGISLATION, REGULATION, AND ETHICAL USE & 
        The development of AI requires legal, regulatory, and ethical parameters to guide its use. It is necessary to find a balance between the protection of rights, encouragement of innovation, and legal security. General principles and ethical parameters should be established through codes of conduct and guidelines, respecting human rights, democratic values, and diversity, with safeguards for human intervention. &
        \begin{itemize}\footnotesize

            \item Encourage the production of ethical AI by funding research projects aimed at applying ethical solutions, especially in the fields of fairness, accountability, and transparency (FAT matrix).
            \item Encourage partnerships with corporations researching commercial solutions for ethical AI technologies.
            \item Establish as a technical requirement in bids that proponents offer solutions compatible with the promotion of ethical AI (e.g., facial recognition technology solutions acquired by public agencies should have a false positive rate below a certain threshold).
            \item Establish, in a multisectoral manner, spaces for discussion and definition of ethical principles to be observed in AI research, development, and use.
            \item Map legal and regulatory barriers to AI development in Brazil and identify aspects of Brazilian legislation that may require updating to promote greater legal security for the digital ecosystem.
            \item Promote transparency and responsible disclosure regarding the use of AI systems and ensure that such systems observe human rights, democratic values, and diversity.
            \item Develop techniques to identify and address algorithmic bias risk.
            \item Develop data quality control policies for training AI systems.
            \item Create parameters for human intervention in AI contexts where the outcome of an automated decision implies a high risk of harm to the individual.
            \item Encourage the exploration and development of appropriate review mechanisms in different contexts of AI use by private organizations and public bodies.
            \item Create and implement best practices or codes of conduct regarding data collection, deployment, and use, encouraging organizations to improve their traceability while safeguarding legal rights.
            \item Promote innovative approaches to regulatory oversight (e.g., sandboxes and regulatory hubs).
        \end{itemize} \\
        \hline
        AI GOVERNANCE & 
        This axis addresses the importance of establishing governance structures to promote the ethical use of AI, emphasizing the need to prevent and eliminate biases in algorithms and databases, as well as the importance of transparency, explainability, and monitoring throughout the AI system lifecycle. Additionally, it highlights the idea of accountability and the participation of various actors in AI development as essential mechanisms for applying the precautionary principle. &
        \begin{itemize}\footnotesize

            \item Structure AI governance ecosystems in the public and private sectors.
            \item Encourage data sharing, in compliance with the General Data Protection Law (LGPD).
            \item Promote the development of voluntary and consensual standards to manage the risks associated with AI applications.
            \item Encourage organizations to create data review boards or ethics committees related to AI.
            \item Create an AI Observatory in Brazil, which can connect to other international observatories.
            \item Encourage the use of representative datasets to train and test models.
            \item Facilitate access to government open data.
            \item Improve the quality of available data to facilitate the detection and correction of algorithmic biases.
            \item Encourage the disclosure of open-source codes capable of verifying discriminatory trends in datasets and machine learning models.
            \item Develop guidelines for the preparation of Data Protection Impact Assessments (DPIA).
            \item Share the benefits of AI development as widely as possible and promote equal development opportunities for different regions and industries.
            \item Develop educational and awareness campaigns.
            \item Encourage social dialogue with multisectoral participation.
            \item Leverage and encourage accountability practices related to AI in organizations.
            \item Define general and specific indicators by sectors (agriculture, finance, health, etc.).
        \end{itemize} \\
        \hline
        INTERNATIONAL ASPECTS & 
        The pursuit of AI leadership requires global cooperation to establish ethical principles, technical standards, and knowledge sharing. Brazil should promote partnerships, actively participate in international forums, and adopt a proactive stance. Cooperation in standards and data protection is essential, aiming at sustainable development and user privacy. &
        \begin{itemize}\footnotesize
            \item Assist the integration of the Brazilian State into international organizations and forums that promote the ethical use of AI.
            \item Promote the exchange of experts developing AI research in various scientific fields, including exact sciences, humanities, and health.
            \item Encourage the export of AI systems developed by Brazilian companies, including startups.
            \item Develop cooperation platforms for information exchange on AI technologies.
        \end{itemize} \\
        \hline
        QUALIFICATIONS FOR A DIGITAL FUTURE & 
        Some countries already offer training in computing and AI, addressing principles and methods. In Brazil, the BNCC highlights the importance of the ethical use of technologies, but qualification for a world with AI involves social and human sciences, in addition to digital skills. AI policies should consider the rights and needs of children, and it is important to avoid excessive monitoring of students. In the job market, digital literacy is essential, as is ethics in the use of information. &
        \begin{itemize}\footnotesize
            \item Evaluate the possibility of updating the BNCC to more clearly incorporate elements related to computational thinking and computer programming.
            \item Develop a digital literacy program in all areas of education and at all education levels.
            \item Expand the offering of undergraduate and graduate courses related to Artificial Intelligence.
            \item Encourage the development of interpersonal and emotional skills, such as creativity and critical thinking (soft skills).
            \item Evaluate ways to incorporate AI technologies in school environments, considering the special condition of children and adolescents as developing individuals and their data protection rights.
            \item Implement technology training programs for teachers and educators.
            \item Include courses on data science basics, linear algebra basics, calculus basics, and probability and statistics basics in the list of complementary activities for high school programs.
            \item Promote interaction programs between the private sector and educational institutions to allow the exchange of practical knowledge about the development and use of Artificial Intelligence technologies.
            \item Create mechanisms to increase Brazilian interest in STEM disciplines (mathematics, science, technology, and engineering) at school age, with a special focus on gender and racial inclusion programs in these areas.
        \end{itemize} \\
        \hline
        WORKFORCE AND TRAINING & 
        Automation and AI will have significant impacts on the job market, and worker retraining, as well as the inclusion of digital skills in the school curriculum, will be important. In Brazil, the number of AI professionals is small compared to the IT sector, and there is low female and racial representation. Thus, public policies should address professional training, general workforce qualification, and the retraining of jobs affected. &
        \begin{itemize}\footnotesize
            \item Establish partnerships with the private sector and academia to define public policies that encourage the training and qualification of professionals, considering the new job market realities.
            \item Encourage companies and public bodies to implement continuous training programs for their workforce focused on new technologies.
            \item Create awareness campaigns on the importance of preparing for the development and ethical use of AI.
            \item Encourage the retention of specialized ICT talent in Brazil.
            \item Encourage the diverse composition of AI development teams in terms of gender, race, sexual orientation, and other sociocultural aspects.
            \item Reinforce policies aimed at continuing education and lifelong learning, promoting greater interaction between the private sector and educational institutions (universities, research institutes, and professional and technical training institutes).
        \end{itemize} \\
        \hline
        RESEARCH, DEVELOPMENT, INNOVATION, AND ENTREPRENEURSHIP & 
        AI will have significant impacts on scientific research and innovation. Governments should promote investments in interdisciplinary R\&D, considering the social, legal, and ethical implications of AI. AI research and development should adopt ethical design approaches, making systems reliable and reducing discrimination. It is necessary to catalyze the execution of AI research and projects and direct funds, such as FNDCT, to AI programs. &
        \begin{itemize}\footnotesize
            \item Define priority areas for AI investments, aligned with other policies related to the digital environment.
            \item Expand the possibilities for research, development, innovation, and application of AI through the allocation of specific resources for this topic and the coordination between existing initiatives.
            \item Establish connections and partnerships between the public sector, private sector, and scientific institutions and universities to advance AI development and utilization in Brazil.
            \item Promote a public policy environment that supports a swift transition from the R\&D phase to the development and operation phase of AI systems.
            \item Promote a research and development environment in AI that is free of bias.
            \item Improve interoperability and the use of common standards.
            \item Promote incentive mechanisms that encourage the development of AI systems that adopt ethical principles and values.
        \end{itemize} \\
        \hline
        APPLICATION IN PRODUCTIVE SECTORS & 
        AI has the potential to make businesses more efficient and reduce operational costs. In Brazil, the pace of AI adoption is increasing, with several areas showing growth potential. Thus, public policies should identify AI usage areas with the best results and prepare the country for the necessary retraining. Investment programs should establish clear objectives and success indicators, considering the impact on fundamental rights. &
        \begin{itemize}\footnotesize
            \item Define or identify a public-private governance structure to promote the advancement of smart IT industries, similar to the Brazilian Industry 4.0 Chamber.
            \item Encourage the emergence of new Brazilian startups in the area through new public-private partnerships.
            \item Create collaboration networks between tech-based startups and small and medium-sized enterprises (SMEs).
            \item Incorporate, in initiatives like the Brazil Mais Program, incentive mechanisms for the use of AI by small and medium-sized enterprises to improve management processes and promote their digital transformation.
        \end{itemize} \\
        \hline
        APPLICATION IN PUBLIC ADMINISTRATION & 
        AI has the potential to improve public services and the quality of citizen care, as well as promote efficient and transparent interaction with citizens and increase trust in the government. In Brazil, public bodies already successfully use AI systems, such as the Federal Court of Accounts (TCU) and the Public Prosecutor's Office. Implementing AI in the public sector requires specific human and organizational skills, as well as considering ethical and transparency issues. &
        \begin{itemize}\footnotesize
            \item In line with the Digital Government Strategy, implement AI resources in at least 12 federal public services by 2022.
            \item Incorporate AI and data analysis in public policy-making processes.
            \item Implement data experimentation spaces with AI and develop AI-focused R\&D partnerships with higher education institutions, the private sector, and the third sector.
            \item Update and reassess processes and work practices in preparation for possible changes in environments where AI systems are introduced.
            \item Consider, in bids and administrative contracts for acquiring AI products and services, criteria not only for technical efficiency but also for incorporating ethical principles related to transparency, fairness, and non-discrimination.
            \item Establish mechanisms for the prompt investigation of complaints and grievances about rights violations in decisions made by AI systems.
            \item Promote the exchange of open data between public administration entities and between them and the private sector, always respecting the right to personal data protection and commercial secrecy.
            \item Conduct impact assessments in cases of AI use that directly affect citizens or public servants.
            \item Establish ethical values for AI use in the Federal Public Administration.
            \item Encourage public bodies to promote awareness of AI use among their technical staff.
        \end{itemize} \\
        \hline
        PUBLIC SECURITY & 
        The application of AI in public security is broad and covers areas such as surveillance, facial recognition, and smart policing; however, concerns about bias, discrimination, and privacy arise. It is crucial that AI technologies respect privacy rights and data protection, and the adoption of technical standards and specific regulations is recommended. &
        \begin{itemize}\footnotesize
            \item Establish supervisory mechanisms for monitoring the use of AI for public security activities.
            \item Encourage agencies that will use AI for monitoring to present data protection impact reports before implementation.
            \item Provide effective mechanisms for monitored individuals to react to surveillance operations.
            \item Present reports with statistics and results of the implemented service.
            \item Develop a law on data protection applied to public security.
            \item Implement a regulatory sandbox for privacy and data protection for AI systems aimed at public security.
        \end{itemize} \\
        \hline
    \end{longtable}

\section*{Declarations}
    \noindent This project is part of the Brazilian Institute of Data Science, supported by FAPESP grant \#2020/09838-0.

    RP received funding from CNPq (grant \#408829/2023-0).

    HSE received funding from FAPESP under the \emph{BRIDGES collaboration} grant \#2021/04065-6 and holds a CNPq Productivity Grant level 1D (grant \#311128/2020-3).

    RS holds a CNPq Productivity Grant level 2 (grant \#311380/2021-2).
    The authors declare no relevant competing interests concerning the content of this article.

\bibliography{refs}


\end{document}